# Exploring the relationship between the Engineering and Physical Sciences and the Health and Life Sciences by advanced bibliometric methods

Ludo Waltman[1], Anthony F.J. van Raan[1], and Sue Smart[2]

[1] Centre for Science and Technology Studies (CWTS), Leiden University, The Netherlands
{waltmanlr, vanraan}@cwts.leidenuniv.nl

[2] Engineering and Physical Sciences Research Council (EPSRC), Swindon, United Kingdom
sue.smart@epsrc.ac.uk

We investigate the extent to which advances in the health and life sciences (HLS) are dependent on research in the engineering and physical sciences (EPS), particularly physics, chemistry, mathematics, and engineering. The analysis combines two different bibliometric approaches. The first approach to analyze the 'EPS-HLS interface' is based on term map visualizations of HLS research fields. We consider 16 clinical fields and five life science fields. On the basis of expert judgment, EPS research in these fields is studied by identifying EPS-related terms in the term maps. In the second approach, a large-scale citation-based network analysis is applied to publications from all fields of science. We work with about 22,000 clusters of publications, each representing a topic in the scientific literature. Citation relations are used to identify topics at the EPS-HLS interface. The two approaches complement each other. The advantages of working with textual data compensate for the limitations of working with citation relations and the other way around. An important advantage of working with textual data is in the in-depth qualitative insights it provides. Working with citation relations, on the other hand, yields many relevant quantitative statistics. We find that EPS research contributes to HLS developments mainly in the following five ways: new materials and their properties; chemical methods for analysis and molecular synthesis; imaging of parts of the body as well as of biomaterial surfaces; medical engineering mainly related to imaging, radiation therapy, signal processing technology, and other medical instrumentation; mathematical and statistical methods for data analysis. In our analysis, about 10% of all EPS and HLS publications are classified as being at the EPS-HLS interface. This percentage has remained more or less constant during the past decade.

## 1. Introduction

During the last two decades, both the UK and the US have, in comparison with other advanced economies, focused a significantly greater proportion of their research budgets on the health and life sciences (HLS). Particularly in the US there has been has been a significant uplift in HLS funding over the last 10 to 20 years (Merrill, 2013). This is despite the fact that it is widely understood and acknowledged that the interdependencies of the various disciplines require that they all advance together. Harold Varmus, former director of the National Institutes of Health in the US, commented that "scientists can wage an effective war on disease only if we – as a nation and as a scientific community – harness the energies of many disciplines, not just biology and medicine". In this paper, we present a detailed analysis of the dependence of HLS advances on research in the engineering and physical sciences (EPS).

Our analysis is of a bibliometric nature, complemented with expert judgment. We analyze large quantities of bibliographic data, both textual data from the titles and abstracts of publications and citation data. The data is taken from Thomson Reuters' Web of Science (WoS) database. Our work can be seen within a long-standing tradition of bibliometric research on interdisciplinarity (e.g., Morillo, Bordons, & Gomez, 2003; Porter & Chubin, 1985; Porter & Rafols, 2009; Van Raan, 2000). The main contribution that we make is in the advanced bibliometric methodology that we use and also in the specific focus of our analysis on interdisciplinary research at the interface between EPS and HLS research fields. We are not aware of earlier studies that have focused specifically on interdisciplinarity at the EPS-HLS interface.

The analysis that we present consists of two parts:

- *Part 1: Analysis based on term map visualizations*. In this analysis, HLS research fields are visualized using so-called term maps (also known as co-word maps), and the role of EPS research in these fields is studied by identifying EPS-related terms in the term maps. The citation impact of the publications in which each term occurs is taken into account as well. The term maps are created using the VOSviewer software (Van Eck & Waltman, 2010, in press).
- *Part 2: Analysis based on a large-scale citation-based clustering of publications*. In this analysis, publications are grouped into clusters based on their citation relations (Waltman & Van Eck, 2012). Each cluster of publications represents a research topic. Citation relations are used to identify topics at the interface between EPS and HLS research fields.

The two parts of the analysis, each based on a very different methodological approach, are intended to be complementary to each other. The advantages of working with textual data (Part 1) compensate for the limitations of working with citation relations (Part 2) and the other way around. An important advantage of working with textual data is in the in-depth qualitative insights it provides. Working with citation relations, on the other hand, yields many relevant quantitative statistics. In cases in which the results of the two parts of our analysis converge, this can be considered to strengthen the overall evidence provided by the analysis.

The organization of this paper is as follows. The methodology and results of Part 1 of our analysis are presented in Section 2. Part 2 of our analysis is discussed in Section 3. Our main conclusions are summarized in Section 4.

## 2. Analysis based on term map visualizations

In this section, we discuss our analysis based on term map visualizations. Term maps are produced for 21 HLS research fields. For each of these fields, the role of EPS research in the field is analyzed by identifying EPS-related terms in the term map of the field. We first present the methodology that we use (Subsection 2.1). We then report the results of the analysis (Subsection 2.2).

### 2.1. Methodology

Our methodology consists of three steps: (1) selection of HLS research fields; (2) production of term maps; and (3) identification of EPS-related terms.

*Step 1: Selection of HLS research fields*

The total number of HLS fields included in the analysis is 21. Of these 21 fields, 16 are in clinical medicine. The other five fields are in the life sciences. The distinction between these two types of fields is important because in comparison with clinical fields, life science fields may be expected to be more closely linked to EPS research.

The 16 clinical fields are listed in Table 1. The five life science fields are listed in Table 2. These 21 fields were chosen to explore the breadth of interactions in areas where EPS was known to have a role. Each field corresponds with a journal subject category in the WoS database. A journal subject category is a set of journals that have been grouped together in the database because they cover the same research field. It is important to be aware that fields are defined at the level of journals, not at the level of individual publications. This for instance means that publications in multidisciplinary journals (e.g., *Nature*, *PLoS ONE*, and *Science*) and in general medical journals (e.g., *Lancet* and *New England Journal of Medicine*) are not included in the 21 fields. Only publications in specialized journals are included. Using our field definitions, a small share of the research in a field, probably with a relatively high citation impact, therefore is excluded from the analysis. We do not expect this to have any systematic effect on the results of the analysis.

Table 1. The 16 clinical fields included in the analysis.

| | |
|---|---|
| cardiac & cardiovascular systems | ophthalmology |
| clinical neurology | orthopedics |
| dentistry | primary health care |
| dermatology | psychiatry |
| hematology | public, environmental & occupational health |
| infectious diseases | respiratory system |
| obstetrics & gynecology | surgery |
| oncology | transplantation |

Table 2. The five life science fields included in the analysis.

| | |
|---|---|
| cell & tissue engineering | materials science, biomaterials |
| chemistry, medicinal | neuroimaging |
| engineering, biomedical | |

*Step 2: Production of term maps*

Visualization is a powerful tool for studying the structure and dynamics of science (Börner, 2010). Many different types of visualizations, often referred to as science maps, are available (Van Eck & Waltman, in press). We use one specific type of visualization, namely term maps. A term map, also known as a co-word map (e.g., Peters & Van Raan, 1993; Van Raan & Tijssen, 1993), is a two-dimensional map of the most important terms in a research field. Frequently occurring terms have a larger size in the map than less frequently occurring terms. Furthermore, terms are positioned in the map in such a way that strongly related terms tend to be located close to each other while less strongly related terms tend to be located further away from each other. In the interpretation of a term map, only the distances between terms are relevant. A map can be freely rotated, because this does not affect the distances between terms. This also implies that the horizontal and vertical axes of a term map have no special meaning.

A term map provides an overview of the structure of a research field. Different areas in a term map correspond with different subfields or different research topics. In the term maps used in our analysis, colors serve to indicate differences in citation impact between subfields (Van Eck, Waltman, Van Raan, Klautz, & Peul, 2013). For each term in a term map, the color of the term is determined by the average citation impact of the publications in which the term occurs. Colors range from blue (low average citation impact) to green (normal average citation impact) to red (high average citation impact). The VOSviewer software (Van Eck & Waltman, 2010, in press) is used to visualize and explore the term maps.

An example of a term map is presented in Figure 1. This is a term map of the *Clinical neurology* field. Each circle represents a term. For some terms, only a circle is displayed, not the term itself. This is done in order to avoid terms from overlapping each other. Using the VOSviewer software, it is possible to zoom in on specific areas in the map. When zooming in, more and more terms will become visible. The term map of the *Clinical neurology* field suggests that the field consists of two subfields (Van Eck et al., 2013). The left area in the map represents the neurosurgery subfield, while the right area represents the neurology subfield. In other words, the left area can be seen as the clinical part of the *Clinical neurology* field, while the right area can be seen as the more basic science part. As can be seen from the coloring of the terms, publications in the neurology subfield on average are cited considerably more frequently than publications in the neurosurgery subfield.

Term maps are produced as follows (Van Eck et al., 2013). First, all publications in a given field in the period 2006–2010 are identified. Only publications classified as article or review in the WoS database are included. For each publication, the number of citations until the end of 2011 is counted. Some fields are larger than others. The number of publications per field therefore ranges between about 5,000 and more than 100,000.

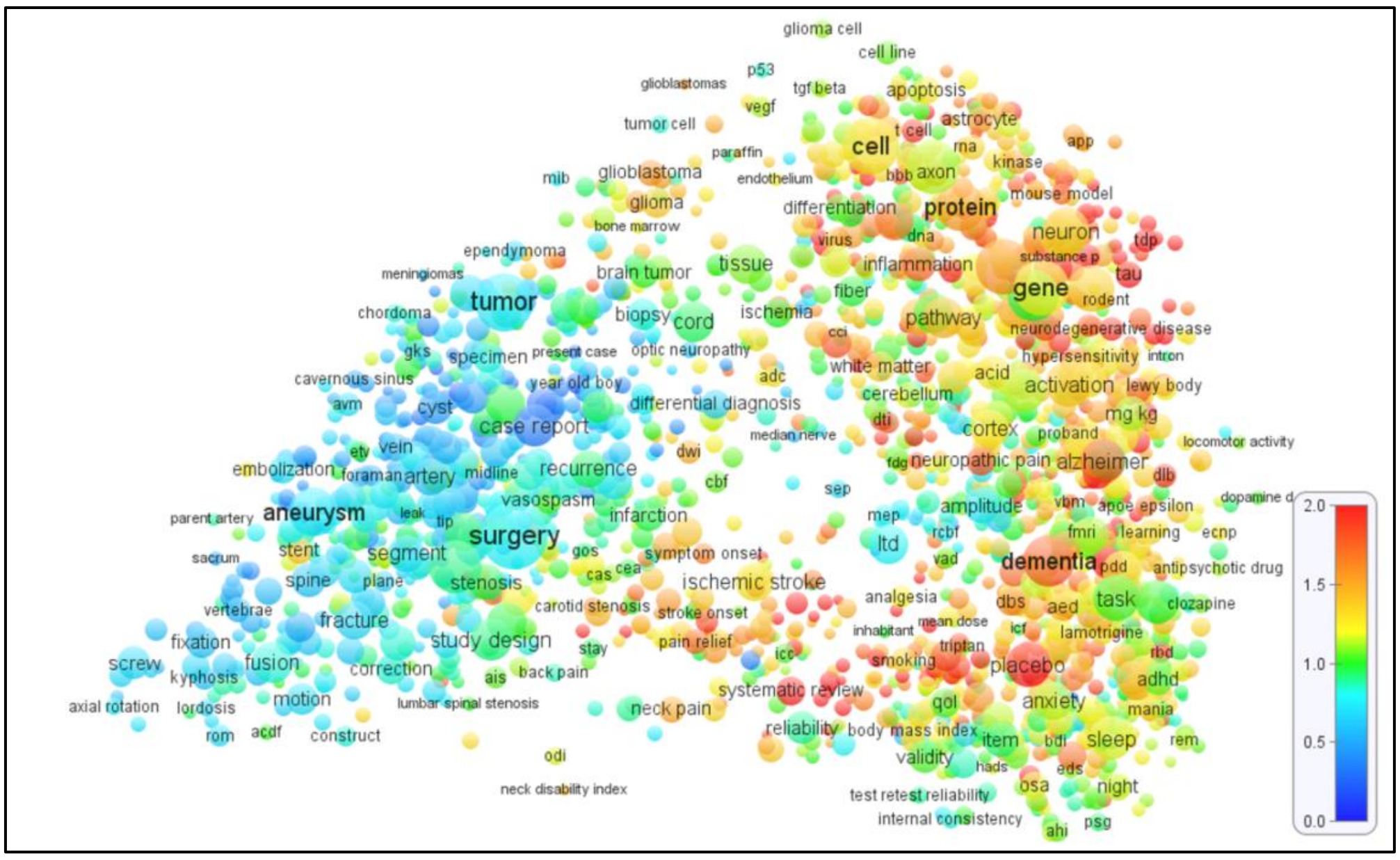

Figure 1. Term map of the *Clinical neurology* field. Colors indicate the average citation impact of the publications in which a term occurs.

Next, using natural language processing techniques, the titles and abstracts of the publications in a field are parsed. This yields a list of all noun phrases (i.e., sequences of nouns and adjectives) that occur in these publications. An additional algorithm (Van Eck & Waltman, 2011) selects 2,000 frequently occurring noun phrases, leaving out general noun phrases such as *result*, *study*, *patient*, and *clinical evidence*. The selected noun phrases can be regarded as the most characteristic terms of a field. Filtering out general noun phrases is crucial. These noun phrases do not relate specifically to a single topic, and they therefore tend to distort the structure of a term map.

Given a selection of 2,000 terms that together characterize a field, the next step is to determine the number of publications in which each pair of terms co-occurs. Two terms are said to co-occur in a publication if they both occur at least once in the title or abstract of the publication. The larger the number of publications in which two terms co-occur, the stronger the terms are considered to be related to each other. The matrix of term co-occurrence frequencies serves as input for the VOS mapping technique (Van Eck, Waltman, Dekker, & Van den Berg, 2010). This technique determines for each term a location in a two-dimensional space. Strongly related terms tend to be located close to each other in the two-dimensional space, while terms that do not have a strong relation are located further away from each other.

In the final step, the color of each term is determined. First, in order to correct for the age of a publication, each publication's number of citations is divided by the average number of citations of all publications that appeared in the same field (i.e., in the same WoS journal subject category) and in the same year. This yields a publication's normalized citation score. A score of 1 means that the number of citations of a publication equals the average of all publications that appeared in the same field and in the same year. Next, for each of the 2,000 terms, the normalized citation scores of all publications in which the term occurs (in the title or abstract) are averaged. The color of a term is determined based on the resulting average score. Colors range from blue (average score of 0) to green (average score of 1) to red (average score of 2 or higher).

*Step 3: Identification of EPS-related terms*

Identification of EPS-related terms is a critical step in our analysis. For each of the 16 clinical fields and the five life science fields, we listed the 2,000 terms included in the term map of the field. The list of terms obtained for each field was inspected manually in order to identify EPS-related terms. The identification of EPS-related terms was carried out by the second author (AFJvR), who has an extensive experience in the natural sciences (especially in experimental and applied physics), in consultation with the third author (SS), who also has significant experience in the natural sciences within both the research and policy contexts.

The focus of the identification procedure is on physics, chemistry, mathematics, and engineering terms. Biology and pharmacy terms are not considered. Typical examples for the *Cardiac & cardiovascular systems* field are 'angiography', 'aortic balloon pump', 'bare metal stent', 'bronchoscopy', 'cardiac magnetic resonance', 'computed tomography', 'confocal microscopy', 'Doppler tissue imaging', 'echocardiography', 'electron microscopy', 'fluorescence', 'high performance liquid chromatography', 'image quality', 'immunohistochemistry', 'intravascular ultrasound', 'mechanical circulatory support', 'optical coherence tomography', 'radiofrequency ablation', 'randomized comparison', and 'statistical model'.

We found for the clinical fields between 50 (*Transplantation*) and 270 (*Dentistry*) EPS-related terms, which corresponds with between 3% and 14% of all terms. For the life science fields, the numbers are mostly considerably higher. Between 207 (*Cell & tissue engineering*) and 817 (*Biomaterials*) of all terms were identified as being EPS-related, which corresponds with between 10% and 41% of all terms. These outcomes are reasonable. Both in the case of the clinical fields and in the case of the life science fields, the field with the largest number of EPS-related terms (*Dentistry* and *Biomaterials*, respectively) is strongly material science oriented and therefore a relatively large number of EPS-related terms can be expected. The percentage of EPS-related terms in each of the 21 fields is reported in Table 3.

Table 3. Percentage of EPS-related terms per field. The 16 clinical fields are listed first, followed by the five life science fields.

| Field | % EPS terms | Field | % EPS terms |
|---|---|---|---|
| cardiac & cardiovas. systems | 4% | psychiatry | 4% |
| clinical neurology | 5% | public, environ. & occup. health | 6% |
| dentistry | 14% | respiratory system | 3% |
| dermatology | 6% | surgery | 5% |
| hematology | 3% | transplantation | 3% |
| infectious diseases | 3% | | |
| obstetrics & gynecology | 5% | cell & tissue engineering | 10% |
| oncology | 10% | chemistry, medicinal | 24% |
| ophthalmology | 6% | engineering, biomedical | 31% |
| orthopedics | 9% | materials science, biomaterials | 41% |
| primary health care | 3% | neuroimaging | 15% |

Terms are often context-dependent. It could be debated whether terms such as ‘air leak’ and ‘capillary’ really do have any engineering relevance. Likewise, chemistry terms such as ‘acetylcholine’ are often names of biochemicals that have a particular significance within the body. Thus such terms are already ‘common property’ in medical research and do not really represent chemical compounds that require specific EPS knowledge to synthesize or detect them. The majority of these cases of doubt concern general physical and, mostly, general chemical terms, particularly chemical compounds. Some examples of such ‘common property’ and/or too general terms are ‘ablation’ (but the more specific term ‘radiofrequency ablation’ is accepted as an EPS-related term), ‘action potential’, ‘adhesion molecule’, ‘air leak’, ‘antioxidant’, chemical formulas such as ‘ca2’, ‘capillary density’, ‘curve analysis’, ‘electrophysiology’, ‘fatty acid’, ‘lipopolysaccharide’, ‘phosphorylation’, ‘polymerase chain reaction’, ‘pulse wave velocity’, and ‘signal’. For all fields, decisions to accept terms as EPS-related have been documented. Undoubtedly, these decisions involve some arbitrariness, but we are confident that at least the accepted terms can be considered as typical EPS terminology.

An additional problem is that some EPS-related terms are in fact synonyms or almost synonyms of each other. An example is given by the terms ‘mri’ and ‘magnetic resonance imaging’. These synonyms may artificially inflate the number of EPS-related terms. This is particularly the case for the *Neuroimaging* field, for which we estimate the influence of synonyms to be about 40% of the EPS-related terms. However, this effect is less problematic than it may seem. Synonyms will generally be located close together in a term map, so they can be easily recognized when inspecting a map.

### 2.2. Results

In this subsection, we provide a detailed discussion of the results obtained for three clinical fields and one life science field. Term maps for all 21 HLS fields are available online at www.cwts.nl/projects/epsrc/. The maps can be explored interactively using the VOSviewer software. In this way, the interested reader can examine our results for each of the 16 clinical fields and each of the five life science fields.

*Clinical fields*

To present the results of our analysis, we use term maps in which EPS-related terms are colored green while all other terms are colored red. These maps are identical to the ones that we use as input for our analysis, except that colors do not reflect citation impact but instead indicate whether a term is considered to be EPS-related or not.

The term map obtained for the *Clinical neurology* field is presented in Figure 2. The map is identical to the one shown in Figure 1, except for the way in which colors are used. In the map in Figure 2, colors do not reflect citation impact but instead indicate the distinction between EPS-related terms (green) and terms not considered to be EPS-related (red). In Figure 3, again the same map is presented, but this time we have zoomed in into the left part of the map around the terms ‘surgery’ and ‘tumor’. This part of the map represents the more clinical part of the *Clinical neurology* field, with a focus on neurosurgical research. We note that inevitably the figures shown in this paper are restricted in size and resolution. To examine the maps in full detail, we recommend to use the online VOSviewer software. The software offers additional functionality, for instance to search for terms or to get information on the number of publications in which a term occurs and on the citation impact of these publications.

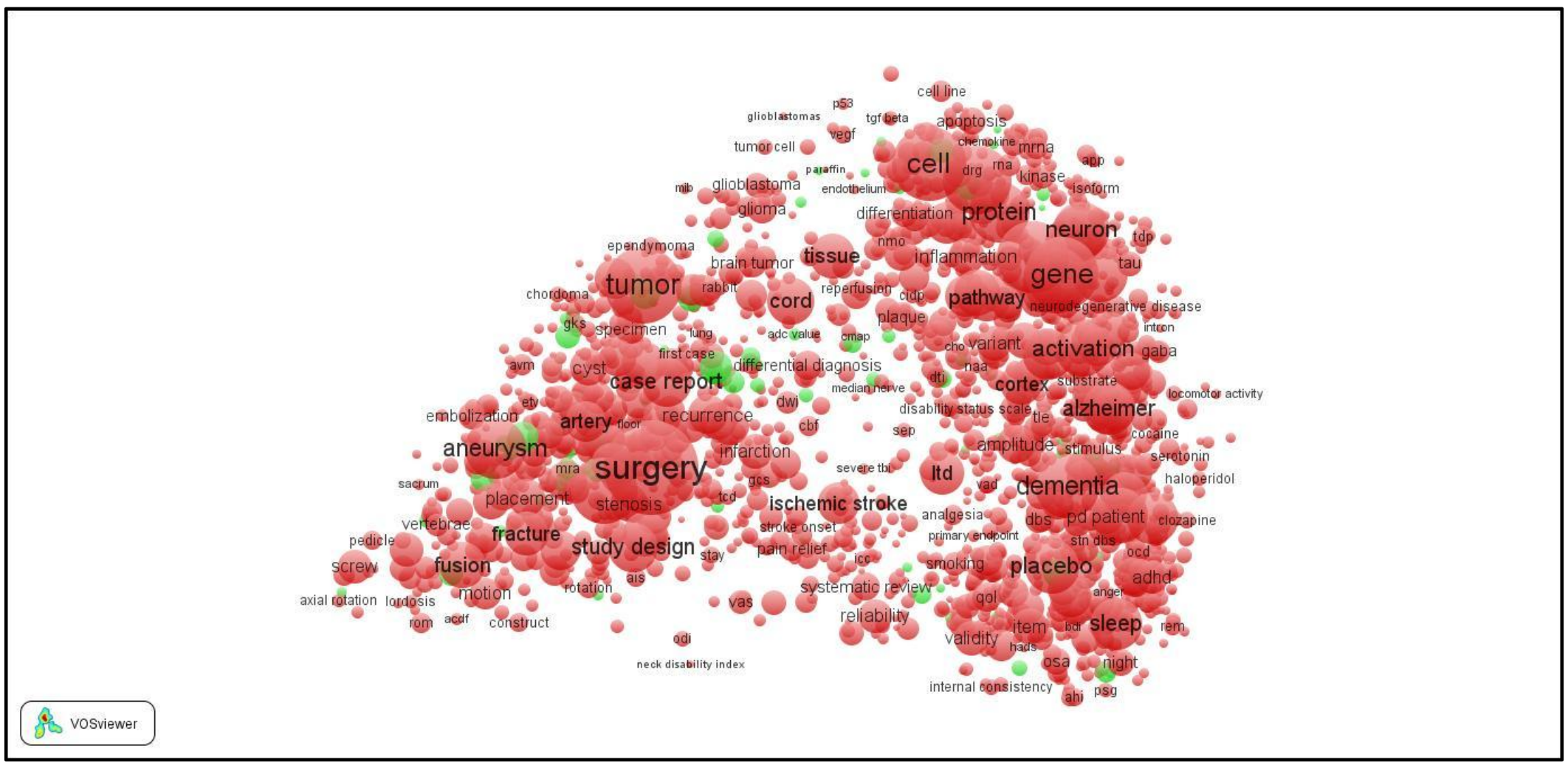

Figure 2. Term map of the *Clinical neurology* field. EPS-related terms are colored green. All other terms are colored red.

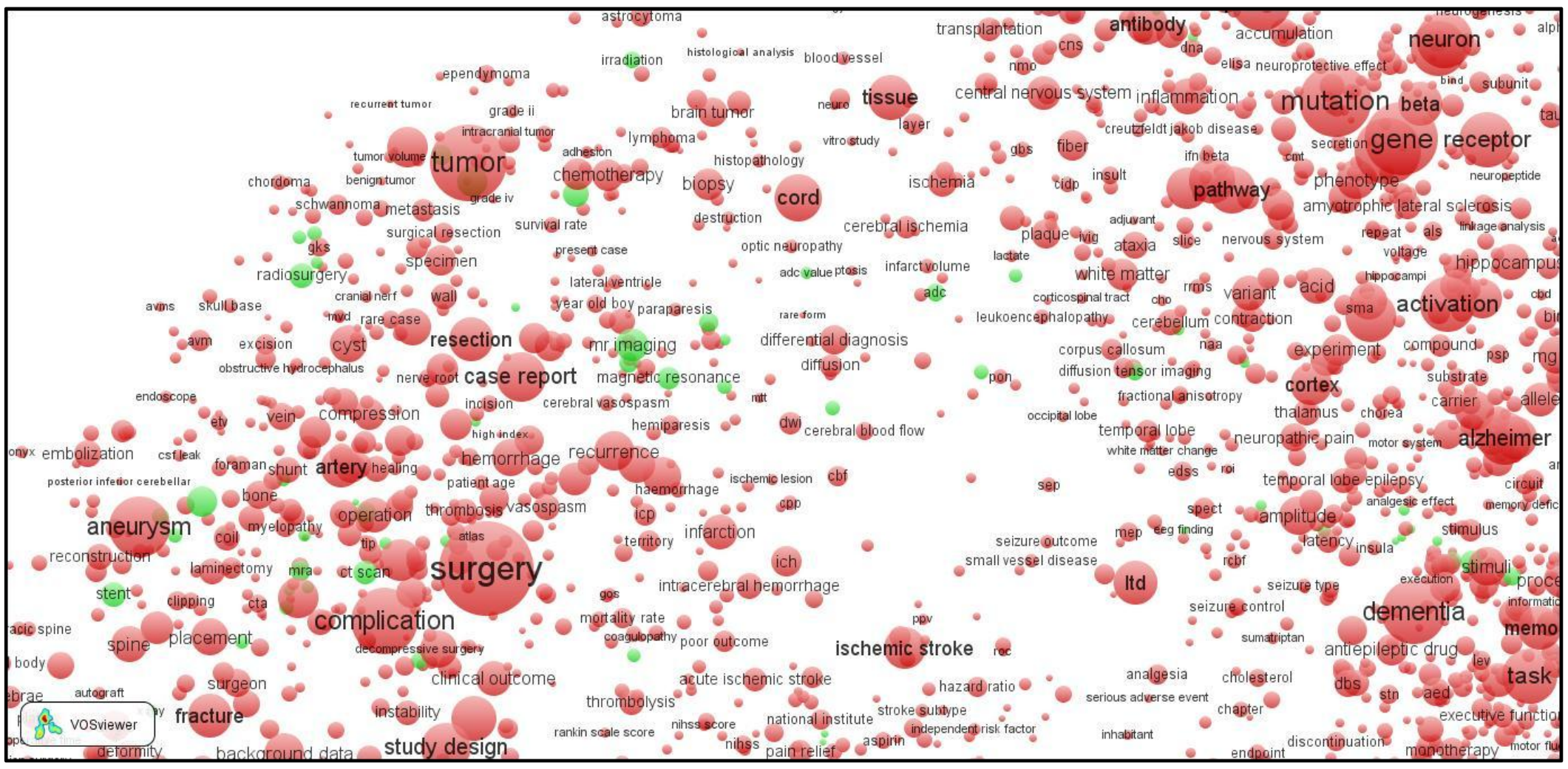

Figure 3. Term map of the *Clinical neurology* field after zooming in into the clinical subfield.

In Figure 3, many EPS-related terms can be found, such as (from top to bottom) ‘radiation’, ‘gamma knife surgery’, ‘radiosurgery’, ‘diffusion weighted imaging’, ‘mr imaging’, ‘magnetic resonance’ (the latter two terms are indeed located close together, as discussed above), ‘angiography’, ‘ct scan’, ‘transcranial Doppler’, ‘X-ray’, etc. These results show that EPS research is prominently present in the ‘hospital side’ (neurosurgery) of the *Clinical neurology* field.

As discussed in Subsection 2.1, the right part of the map of the *Clinical neurology* field represents the more basic science part of the field. In Figure 4, we have zoomed in into this part of the map, around the terms ‘cell’ and ‘expression’. Instead of hospital-related terms such as ‘imaging’, ‘radiation’, and ‘mr spectroscopy’, we now find EPS-related terms corresponding to typical basic science topics. Some examples of such terms are ‘immunohistochemical analysis’, ‘molecular analysis’, ‘molecular basis’, ‘molecular mechanism’, etc.

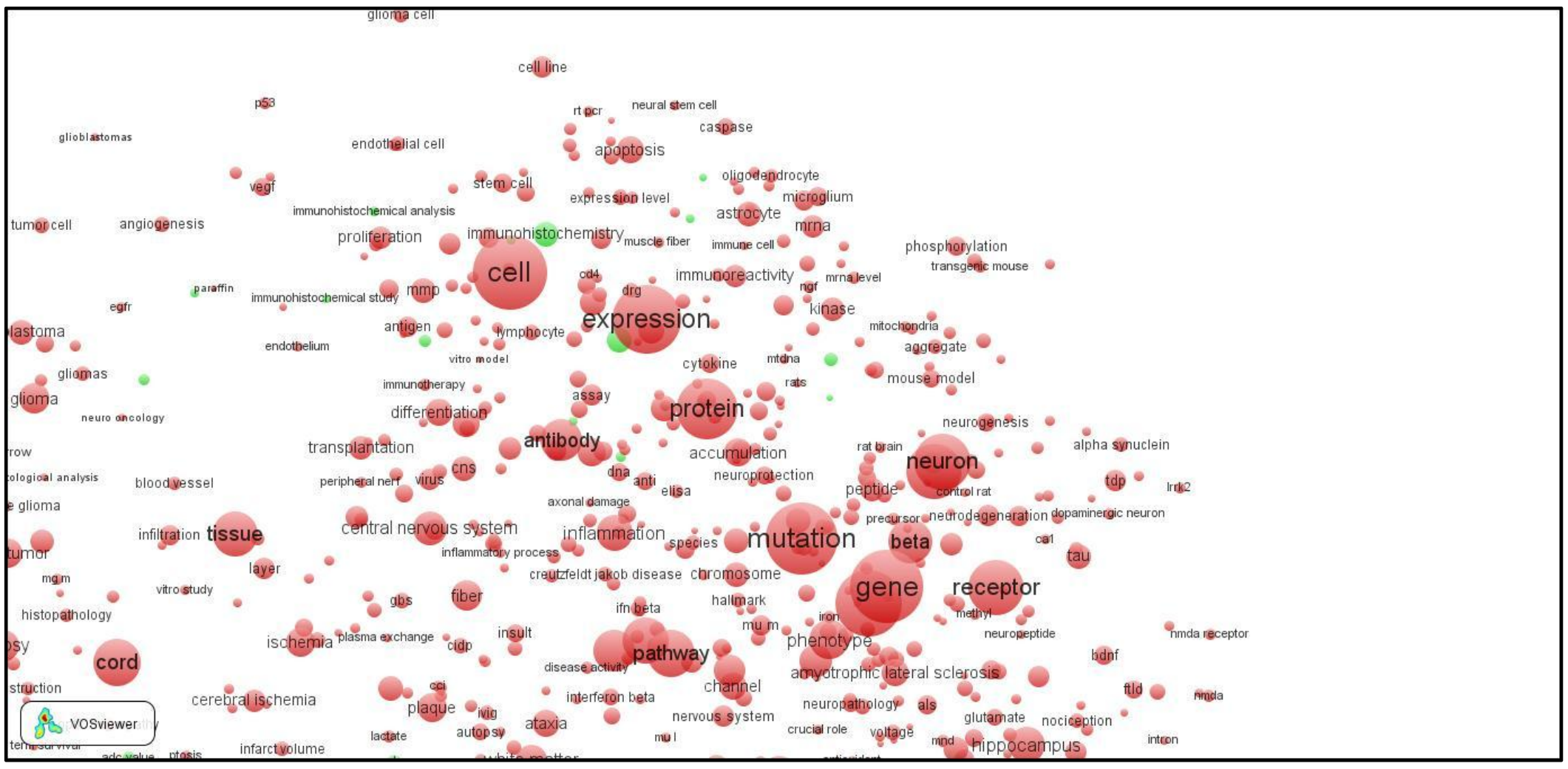

Figure 4. Term map of the *Clinical neurology* field after zooming in into the basic science subfield.

Using the VOSviewer software, it is also possible to identify EPS-related topics with a high citation impact in the term maps. Examples of high-impact EPS-related topics in the *Clinical neurology* field include research on stents, research related to neuroimaging (terms such as ‘diffusion tensor imaging’, ‘diffusion weighted imaging’, ‘functional magnetic resonance imaging’, and ‘tractography’), sleep research (‘actigraphy’), immunocytochemistry, and statistical methods (‘disease rating scales’, ‘double blind studies’, ‘multivariable analysis’, and ‘randomization’).

As can be seen in Table 3, *Clinical neurology* is a field with a relatively low percentage of EPS-related terms (5%). *Dentistry* is the clinical field with the highest percentage of EPS-related terms (14%). This is clearly visible in the *Dentistry* term map presented Figure 5. We observe that particularly the left part of the map, corresponding to the more clinically oriented (hospital) subfield of the *Dentistry* field, is dominated by EPS-related terms. When zooming in into this part of the map (see Figure 6), we observe that these EPS-related terms represent mainly dental materials, as can be expected. Examples of EPS-related terms in this part of the map include ‘adhesive resin’, ‘alloy’, ‘bond strength’, ‘cement’, ‘ceramic’, ‘coating’, ‘composite’, ‘metal’, ‘porcelain’, and ‘powder’.

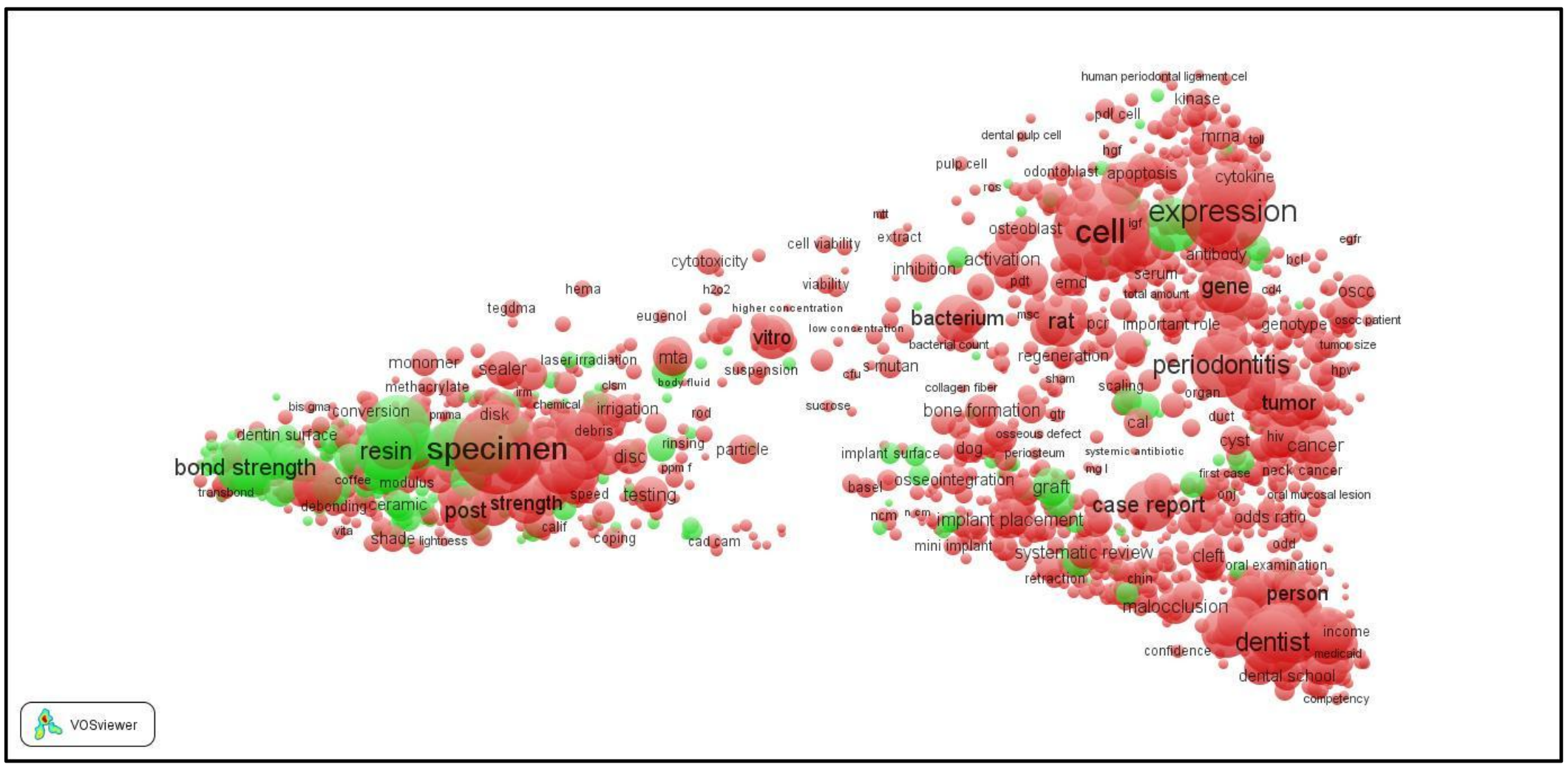

Figure 5. Term map of the *Dentistry* field.

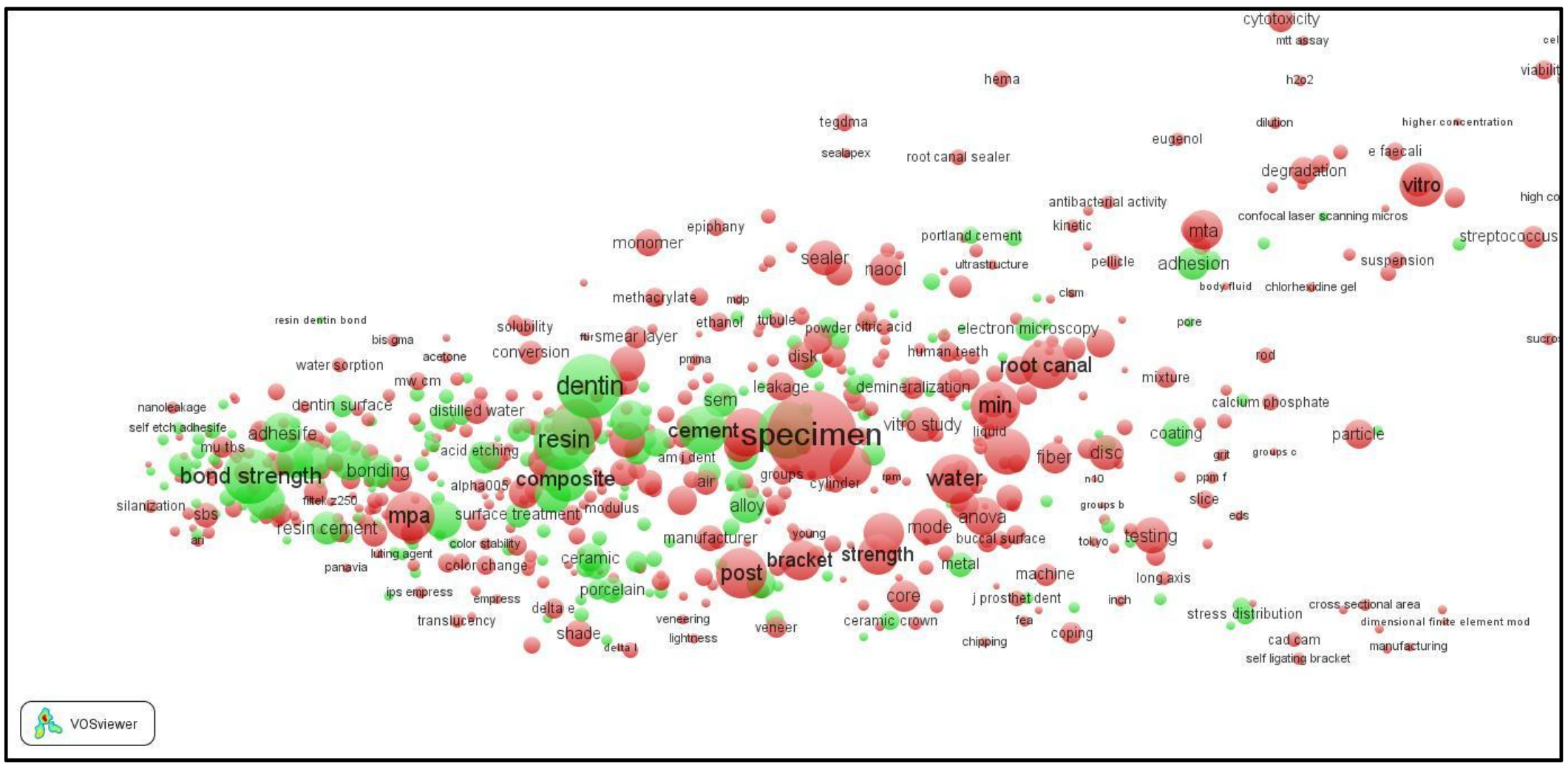

Figure 6. Term map of the *Dentistry* field after zooming in into the clinical subfield.

Many of the EPS-related terms concern high-impact research. Without exaggeration, one can say that the clinical subfield of the *Dentistry* field is to a large extent driven by high-impact EPS research. This high-impact EPS research relates to dental materials science work that can be divided into research on materials (terms such as 'bond strength', 'ceramics', 'coating', 'cohesive failures', 'compressive strength', 'elasticity', 'grafting material', 'implant stability', 'phosphoric acid etching', 'polymerization', 'Portland cement', 'powder', and 'zirconia') and, particularly, materials surface research ('atomic force microscopy', 'confocal laser scanning microscopy', 'fracture surface', 'scanning electron microscopy', and 'transmission electron microscopy'). High-impact EPS research also concerns imaging technology, with terms such as 'cephalometric radiography', 'cone beam computed tomography', and 'X-ray diffraction'.

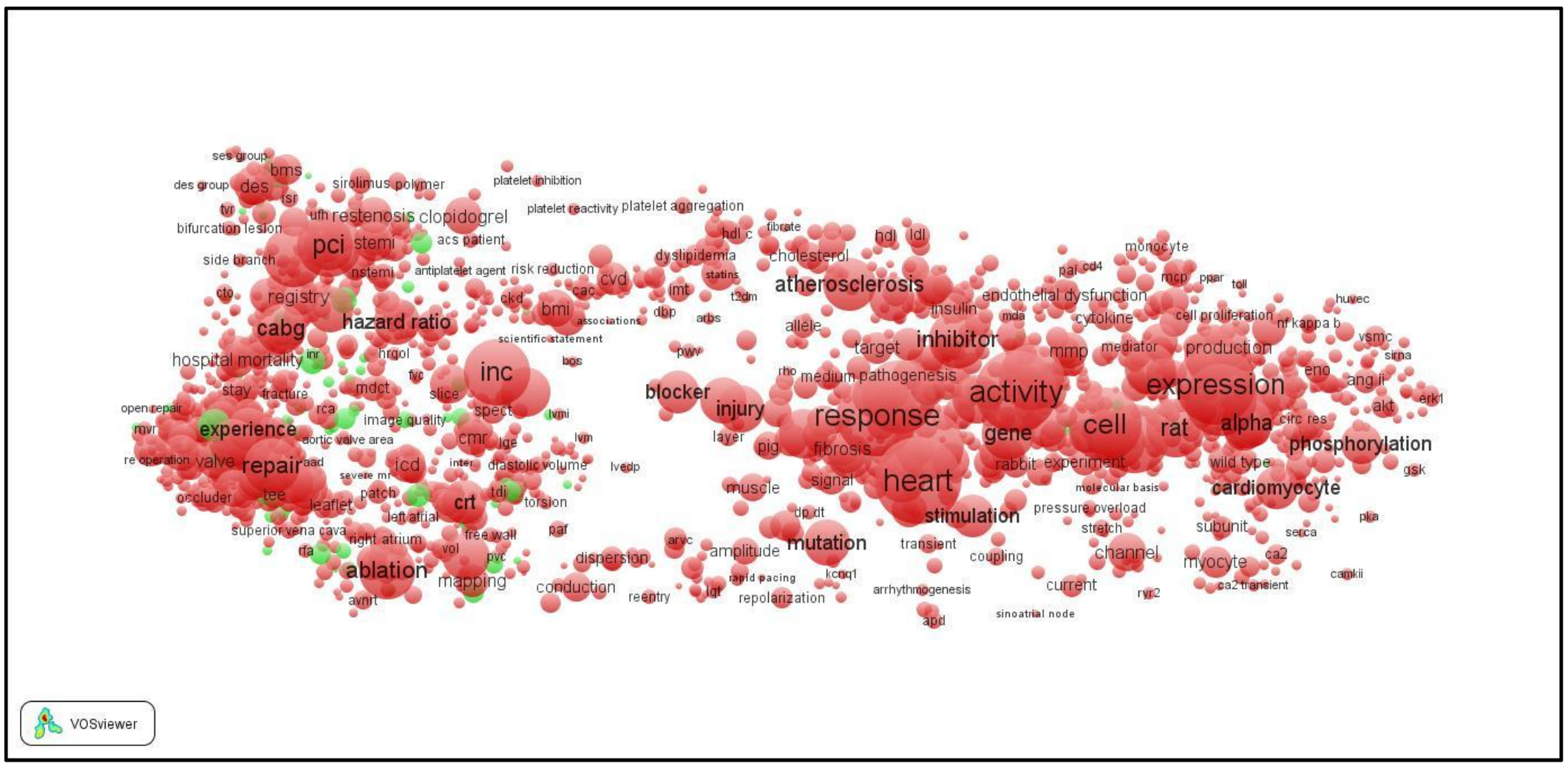

Figure 7. Term map of the *Cardiac & cardiovascular systems* field.

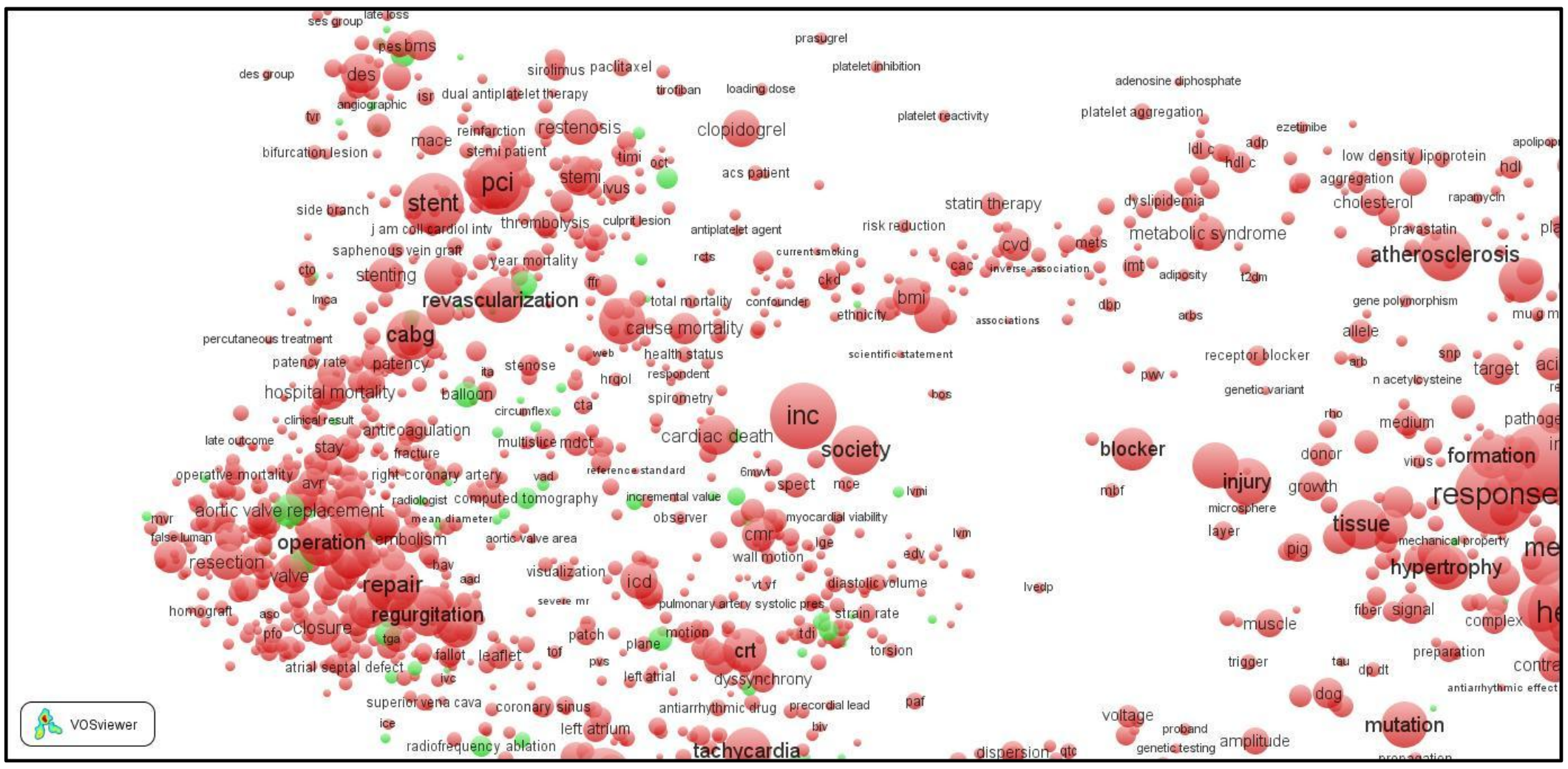

Figure 8. Term map of the *Cardiac & cardiovascular systems* field after zooming in into the clinical subfield.

As a third and last example of the 16 clinical fields, we show in Figure 7 the term map of the *Cardiac & cardiovascular systems* field. Here we also observe some intriguing features. The majority of the EPS-related terms is located in left part of the map, which again represents the hospital/clinical subfield. Zooming in into this part of the map (see Figure 8) reveals many EPS-related terms: ‘bare metal stent’, ‘computed tomography’, ‘echocardiography’, ‘fluoroscopy’, ‘intravascular ultrasound’, ‘radiofrequency ablation’, ‘tissue Doppler imaging’, etc.

Also in the *Cardiac & cardiovascular systems* field, we find high-impact EPS research. This field is a good example of a general observation: high-impact EPS work in clinical fields often concerns (1) new materials and their properties (in this field terms such as ‘bare metal stents’ and ‘stent fractures’), (2) chemical methods (‘high performance liquid chromatography’, ‘immunocytochemistry’, and ‘immunofluorescence’), (3) imaging (‘confocal microscopy’, ‘echocardiography’, ‘intravascular ultrasound’, ‘invasive coronary angiography’, and ‘optical coherence tomography’), (4) medical engineering (‘transcatheter aortic valve implantation’), and (5) mathematical and statistical methods (‘randomized trial’).

As already mentioned, readers are invited to use the VOSviewer software for higher resolution analyses and for exploring other clinical fields.

*Life science fields*

We now consider the life science fields. We focus on one field, *Biomedical engineering*. In Table 3, we see that this field has a high percentage of EPS-related terms (31%), as can be expected for a field with ‘engineering’ in its name. The term map of the *Biomedical engineering* field is shown in Figure 9. We observe the remarkably ‘polarized’ structure of the field. It is as if the field falls apart into two more or less separated subfields. The left part of the map is dominated by materials science and the right part by imaging and radiotherapy techniques. Nevertheless, the connection between these two subfields is very well understandable. Many of the imaging techniques are necessary to study the surface properties of new biomaterials. When we zoom in into the right part of the map, we find a recent development in the treatment of cancer, proton therapy. This development is indicated by the term ‘proton beam’ in the bottom part in Figure 10. Although this term does not appear very prominently in the map (it occurs in 89 publications), it is clearly embedded in an area characterized by radiotherapeutical techniques.

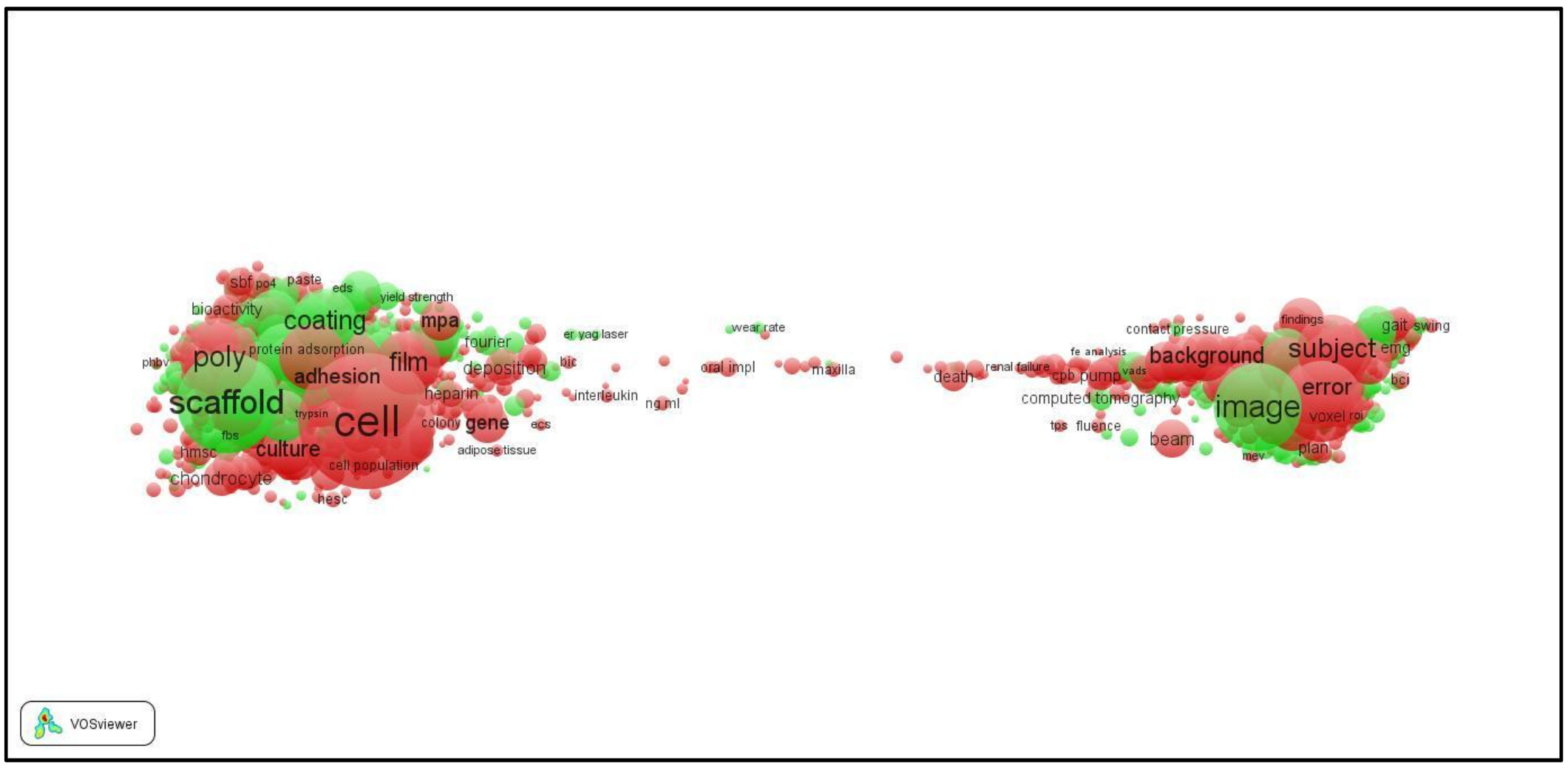

Figure 9. Term map of the *Biomedical engineering* field.

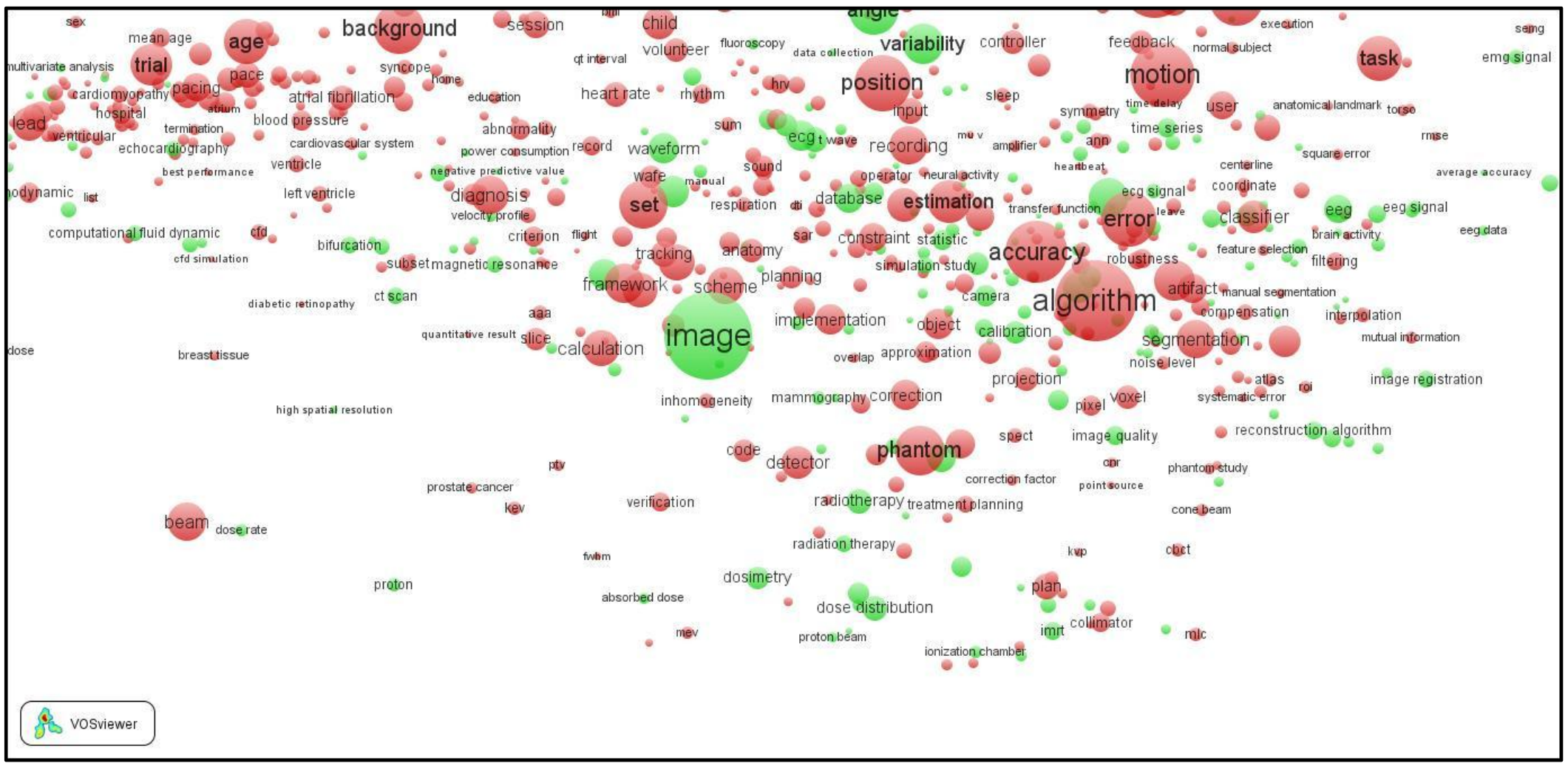

Figure 10. Term map of the *Biomedical engineering* field after zooming in into the imaging and radiation therapy subfield.

As can be expected, the life science field *Biomedical engineering* is very strongly EPS-driven. Many of the EPS-related terms concern physics, chemistry, and engineering research of high to very high impact. Again, we find the main EPS areas mentioned earlier in the context of the clinical fields, that is, materials science, chemistry, imaging, engineering, and statistics. In the *Biomedical engineering* field, a particularly strong focus is on the development of new biomaterials, which combines research in physics, chemistry, and engineering. Examples of terms are 'bone-tissue regeneration', 'cartilages', 'cell-material interaction', 'composite materials', 'polymers', 'scaffolds', and 'self-assembly'. Most of the imaging work relates to the study of the surfaces of new biomaterials, with terms such as 'confocal laser scanning microscopy', 'dynamic light scattering', 'fluorescence microscopy', 'nmr', 'transmission electron microscopy', and 'X-ray diffraction'. Examples of terms indicating high-impact chemical research are 'cytotoxicity', 'immunohistochemistry', 'lactic acid', 'model proteins', and 'surface chemistry'.

The very high impact of research on the construction of nanoparticles and nanocomposites and particularly research on electrospinning (i.e., a method to draw very fine fibers from a liquid, typically on the micro- or nanoscale) and on quantum dots (i.e., nanocrystals made of semiconductor materials that are small enough to exhibit quantum mechanical properties) is remarkable. The terms 'electrospinning' and 'quantum dot' can be found by using the VOSviewer software in the left part of the map, close to the term 'scaffold'.

## 3. Analysis based on citation-based clustering of publications

This section focuses on our analysis based on a large-scale citation-based clustering of publications. In this analysis, we first use citation relations to group publications into clusters. Each cluster of publications represents a research topic. We then use citation relations to identify topics that are located at the interface between EPS and HLS research fields. We are particularly interested in topics in EPS research fields that, based on citation patterns, seem to have a strong influence on HLS research fields.

We first discuss our methodology (Subsection 3.1), and we then present the results of our analysis (Subsection 3.2).

### 3.1. Methodology

The methodology that we use consists of three steps: (1) clustering of publications based on citation relations; (2) identification of research topics at the EPS-HLS interface; and (3) aggregation of EPS-HLS research topics into broad research themes. We now discuss the above steps one by one.

*Step 1: Clustering of publications based on citation relations*

We start by constructing a large-scale clustering of publications based on citation relations. This step is discussed in detail in an earlier paper (Waltman & Van Eck, 2012). Here we summarize the main elements of the approach that we follow. We refer to Boyack and Klavans (2014) for an alternative approach to the large-scale citation-based clustering of publications.

We take all 10.2 million publications in the WoS database in the period 2001–2010, and we collect all 97.6 million citation relations between these publications. Based on these citation relations, we group closely connected publications together into clusters. This is done using a clustering technique. The overall number of clusters that we obtain is 22,412. Each cluster includes at least 50 and at most about 4,000 publications, with an average of 422 publications per cluster. Each cluster can be interpreted as a research topic in the scientific literature. The 22,412 topics cover all scientific disciplines, including the social sciences. Some topics are highly interdisciplinary and cover publications from many different research fields.

Each of the 22,412 topics is labeled in an algorithmic way. This is done by extracting the most relevant terms from the titles and abstracts of the publications belonging to a topic. For each topic, five terms are selected. Terms are selected based on two criteria. On the one hand, terms must be of sufficient importance, which means that they must occur in a sufficiently large number of publications. On the other hand, terms must be sufficiently unique. In other words, in order to properly characterize a topic, terms should not be too general. We therefore leave out terms that relate to many different topics.

Our clustering of 10 million publications into 22,412 topics offers a unique and highly detailed structure of science. This structure is much more detailed than for instance the structure provided by the WoS journal subject categories used in Section 2. Moreover, the structure is not only more detailed, but it can also be expected to be significantly more accurate, since the structure is created at the level of individual publications rather than at the level of entire journals. This means that especially publications in large journals with a broad scope and publications in multidisciplinary journals such as *Nature*, *PLoS ONE*, and *Science* can be handled in a more accurate way.

*Step 2: Identification of research topics at the EPS-HLS interface*

The next step is to select among the 22,412 topics identified in step 1 the topics that are at the interface between EPS and HLS research fields. More precisely, our aim is to select topics that include a significant share of EPS publications while at the same time they receive a significant share of their citations from HLS publications. These topics can be expected to represent EPS research that has a strong influence on HLS research.

To identify topics that are at the interface between EPS and HLS research fields, we first need to define what we consider to be EPS and HLS research fields. Although as discussed above we regard the WoS journal subject categories as rather crude structures, we can use them for this purpose. There are about 250 subject categories, representing research fields in the sciences, the social sciences, and the arts and humanities. We have selected 72 subject categories as EPS research fields. These fields are listed in Table A1 in the appendix. 86 subject categories, listed in Table A2 in the appendix, have been selected as HLS research fields.

Based on our selection of EPS and HLS research fields, we calculate for each of the 22,412 topics identified in step 1 the percentage of publications in journals in EPS fields. We also calculate for each topic the percentage of citations received from journals in HLS fields. A topic is considered to be at the EPS-HLS interface if two conditions are met. On the one hand, the topic must have at least a certain minimum percentage of EPS publications. On the other hand, at least a certain minimum percentage of the citations received by the publications belonging to the topic must originate from HLS publications. For both criteria, a threshold of 34% (i.e., roughly one third) has been chosen. The choice of this threshold of course involves some arbitrariness. It is based on an analysis which suggests that on the one hand a threshold of 34% ensures that the most important topics at the EPS-HLS interface are all included while on the other hand we still remain reasonably selective in what we consider to be research at the EPS-HLS interface. The use of a threshold of 34% results in the selection of 959 topics at the EPS-HLS interface. Choosing a somewhat higher or lower threshold would have led to a somewhat smaller or larger number of topics, but the results are reasonably robust to the choice of the threshold.

*Step 3: Aggregation of EPS-HLS research topics into broad research themes*

To facilitate further analysis, the 959 research topics at the EPS-HLS interface identified in step 2 are grouped into a limited number of broad research themes. This is done based on citation relations between publications belonging to the different topics. Like in step 1, a clustering technique is used, although some manual adjustments are made as well. We obtain 11 research themes. Each theme is given a label. This is done manually, based on an examination of the research covered by each theme.

### 3.2. Results

Below, we first discuss the results obtained at the level of the 959 research topics identified at the EPS-HLS interface (step 2 in Subsection 3.1). We then discuss the results obtained at the level of 11 broad research themes (step 3 in Subsection 3.1). Finally, we illustrate how our results can be used to provide information on the contribution of for instance countries or research institutions to research at the EPS-HLS interface.

*Research topics at the EPS-HLS interface*

The 959 research topics identified at the EPS-HLS interface include 862,565 publications in the period 2001–2010. In the same period, the total number of publications in the WoS database (articles and reviews only) in EPS fields is about 3.77 million. The total number of publications in HLS fields is about 4.35 million. Hence, based on the criteria discussed in Subsection 3.1, about 0.86 / (3.77 + 4.35) = 10.6% of all EPS and HLS publications are considered to be at the interface between EPS and HLS fields. For the period 2001–2010, no evidence was found for either an increasing or a decreasing trend in the percentage of publications at the EPS-HLS interface.

To illustrate the types of research topics identified at the EPS-HLS interface, we focus on topics that have experienced a strong growth in publication output during the period 2001–2010. These topics may be considered emerging topics. Of the 959 research topics at the EPS-HLS interface, Table 4 lists the 20 topics with the most significant growth in publication output.[1] As explained in Subsection 3.1, each topic

[1] These 20 topics satisfy the following criteria: (1) the number of publications in 2010 is at least four times as large as the number of publications in 2001; (2) the number of publications in 2001 is at most

is labeled using a number of terms that have been algorithmically identified and that are expected to provide a good indication of what the topic is about. For each topic, Table 4 also lists the number of publications as well as the broad research theme to which the topic has been assigned. We will get back to these research themes below.

Table 4. The 20 research topics at the EPS-HLS interface with the most significant growth in publication output. The number of publications relates to the period 2001–2010.

| Research topic | No. of pub. | Research theme |
|---|---|---|
| microenvironment; culture; perfusion; cellular response; chemotaxis | 1194 | Biological analysis |
| high content screening (hcs); segmentation; image data; rna interference (rnai) | 696 | Biological analysis |
| emission depletion; diffraction barrier; lateral resolution; point spread function (psf) | 696 | Biological analysis |
| tissue section; imaging mass spectrometry; matrix-assisted laser desorption ionization (maldi) imaging; spatial distribution; tissue surface | 383 | Biological analysis |
| carbon nanotube; nanomaterial; nanotechnology; multi-wall carbon nanotubes (mwcnt); titanium dioxide (tio2) | 1663 | Biomedical engineering and brain/neural |
| microbial fuel cell (mfc); anode; electricity | 1039 | Biomedical engineering and brain/neural |
| brain computer interface (bci); bci system; motor imagery; mental task | 812 | Biomedical engineering and brain/neural |
| critical assessment of prediction of interactions (capri); protein-protein docking; interface residue; protein interface; hot spot | 743 | Genomics and proteomics |
| elastic network model (enm); normal mode analysis; adenylate kinase; allostery | 726 | Genomics and proteomics |
| ligand binding site; catalytic residue; pocket; functional site; unknown function | 685 | Genomics and proteomics |
| solid lipid nanoparticle (sln); nanostructured lipid carrier (nlc); lipid matrix | 629 | Materials for drug delivery and controlled release |
| feature selection; cancer classification; support | 1011 | Medical statistics and |

30; and (3) the number of publications in 2010 is at least 60. For an alternative approach to identifying emerging topics, see Small, Boyack, and Klavans (in press).

| | | |
|---|---|---|
| vector machine (svm); classifier | | informatics |
| reverse engineering; bayesian network; microarray data; gene expression data; regulatory relationship | 971 | Medical statistics and informatics |
| extracellular signal-regulated kinases (erk); mitogen-activated protein kinase (mapk); epidermal growth factor (egf) receptor; receptor | 722 | Medical statistics and informatics |
| gene ontology; go term; gene set enrichment analysis (gsea); gene set; enrichment analysis | 614 | Medical statistics and informatics |
| structure activity relationship (sar); hepatitis c virus nonstructural protein 5b (hcv ns5b) polymerase; boceprevir; compound | 822 | Medicinal chemistry |
| microtubule-associated protein 2 (mk2); map kinase inhibitor; fluorophenyl; alpha map kinase; birb | 648 | Medicinal chemistry |
| linoleic acid emulsion; superoxide anion radical scavenging; hydrogen peroxide scavenging; metal chelating activity; standard antioxidant | 350 | Natural products for pharmaceutical use |
| bacterium; photodynamic inactivation; staphylococcus aureus; biofilm; escherichia coli | 602 | Pharmaceutical and food analysis |
| melamine; cyanuric acid; milk; pet food; milk powder | 318 | Pharmaceutical and food analysis |

*Broad research themes at the EPS-HLS interface*

The 959 research topics at the EPS-HLS interface have been grouped into 11 broad research themes. This was done mostly in an algorithmic way based on citation relations between publications belonging to the different topics, but some manual work was done as well. First, the 959 research topics were algorithmically grouped into 21 clusters. Next, clusters that we considered to be strongly related to each other were merged. This resulted in 11 broad research themes. Of the 11 broad research themes, some have a somewhat heterogeneous nature. This is in particular the case for the research theme labeled *Biomedical engineering and brain/neural*. We have considered splitting up this theme into two separate themes, but there turned out to be relatively strong citation relations between the research topics included in the theme, making it difficult to split up the theme in a satisfactory way.

The 11 broad research themes are listed in Table 5. For each research theme, the table indicates the number of publications as a percentage of the total number of publications at the EPS-HLS interface (i.e., the total numbers of publications included in the 959 research topics). The largest research theme, in terms of its number of publications, is *Biomedical engineering and brain/neural*. This theme includes 13.4% of all publications at the EPS-HLS interface. The smallest research themes, each with 6.0% of all publications at the EPS-HLS interface, are *Food chemistry* and *Materials for drug delivery and controlled release*.

Table 5. The 11 broad research themes at the EPS-HLS interface. For each research theme, the number of publications as a percentage of the total number of publications at the EPS-HLS interface is reported in the right column.

| Research theme | % of pub. |
|---|---:|
| Biological analysis | 6.7% |
| Biomaterials | 7.6% |
| Biomedical engineering and brain/neural | 13.4% |
| Food chemistry | 6.0% |
| Genomics and proteomics | 8.4% |
| Materials for drug delivery and controlled release | 6.0% |
| Medical imaging and radiotherapy | 8.0% |
| Medical statistics and informatics | 7.3% |
| Medicinal chemistry | 13.2% |
| Natural products for pharmaceutical use | 12.4% |
| Pharmaceutical and food analysis | 11.0% |

A visual representation of the 11 broad research themes and their location within the general structure of science is presented in Figure 11. Each dot in this figure represents one of the 22,412 research topics identified in step 1 of our methodology. The dots have been positioned algorithmically in such a way that research topics that are strongly connected to each other by citation relations tend to be located close to each other in the figure. Labels have been manually added to the figure to roughly indicate the locations of a number of broad scientific disciplines. The colored dots in Figure 11 represent the 959 research topics at the EPS-HLS interface identified in step 2 of our methodology. The color of a dot indicates to which of the 11 broad research themes a topic has been assigned in step 3 of our methodology.

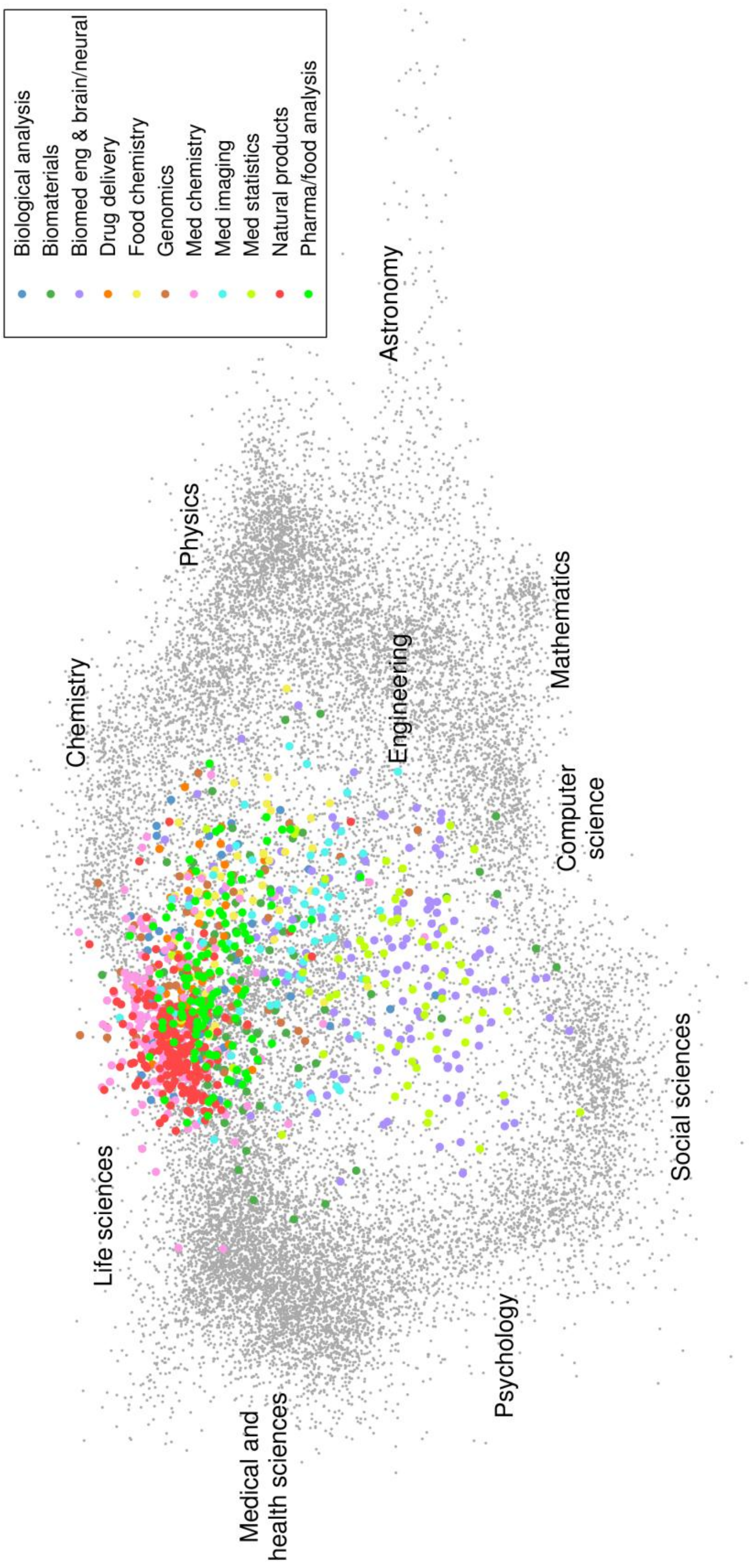

Figure 11. The 11 broad research themes at the EPS-HLS interface and their location within the general structure of science.

The visualization presented in Figure 11 shows a kind of circular structure. This typical structure of science has also been found in various earlier studies (e.g.,

Klavans & Boyack, 2009; Van Eck & Waltman, 2010). Moving in clockwise direction and starting in the left part of the visualization, we first observe the medical and health sciences and the life sciences, followed by chemistry, physics, astronomy,[2] engineering, mathematics, and computer science. Computer science, in turn, is close to the social sciences, the social sciences are close to psychology, and finally the circle is completed by the close relationship between psychology and the medical and health sciences.

As expected, the colored dots in Figure 11, representing the 11 broad research themes at the EPS-HLS interface, are mainly located in between the EPS and HLS research fields, but there are also dots that are located close to psychology and the social sciences. Furthermore, some of the research themes seem to be quite concentrated in a relatively small part of science, while other themes seem to have a much more interdisciplinary nature. For instance, the *Drug delivery* and *Natural products* themes are located mainly around the life sciences in Figure 11, while the *Biomaterials, Biomedical engineering and brain/neural*, and *Medical imaging* themes clearly have relations to physics, engineering, and computer science. The *Medical statistics* theme is connected to psychology and the social sciences, which is understandable given the similarity in the statistical techniques that are used.

In Figure 12, we show for each of the 11 broad research themes at the EPS-HLS interface how the number of publications has evolved between 2001 and 2010. An increasing trend can be observed for all 11 research themes. In itself, this increasing trend is of limited interest. Similar trends can be observed for most fields of science. This is a consequence of on the one hand the increasing number of publications that appear each year in the scientific literature and on the other hand the increasing coverage of the WoS database that we use in our analysis (i.e., journals previously not covered by the WoS database are added to the database). In fact, a further analysis reveals that in the period 2001–2010 the number of publications belonging to the 11 research themes at the EPS-HLS interface has grown at the same rate as the overall number of publications in the WoS database.

[2] Notice the peripheral position of astronomy in Figure 11. In fact, some research topics in astronomy are even more peripheral and therefore have not been included at all in the figure.

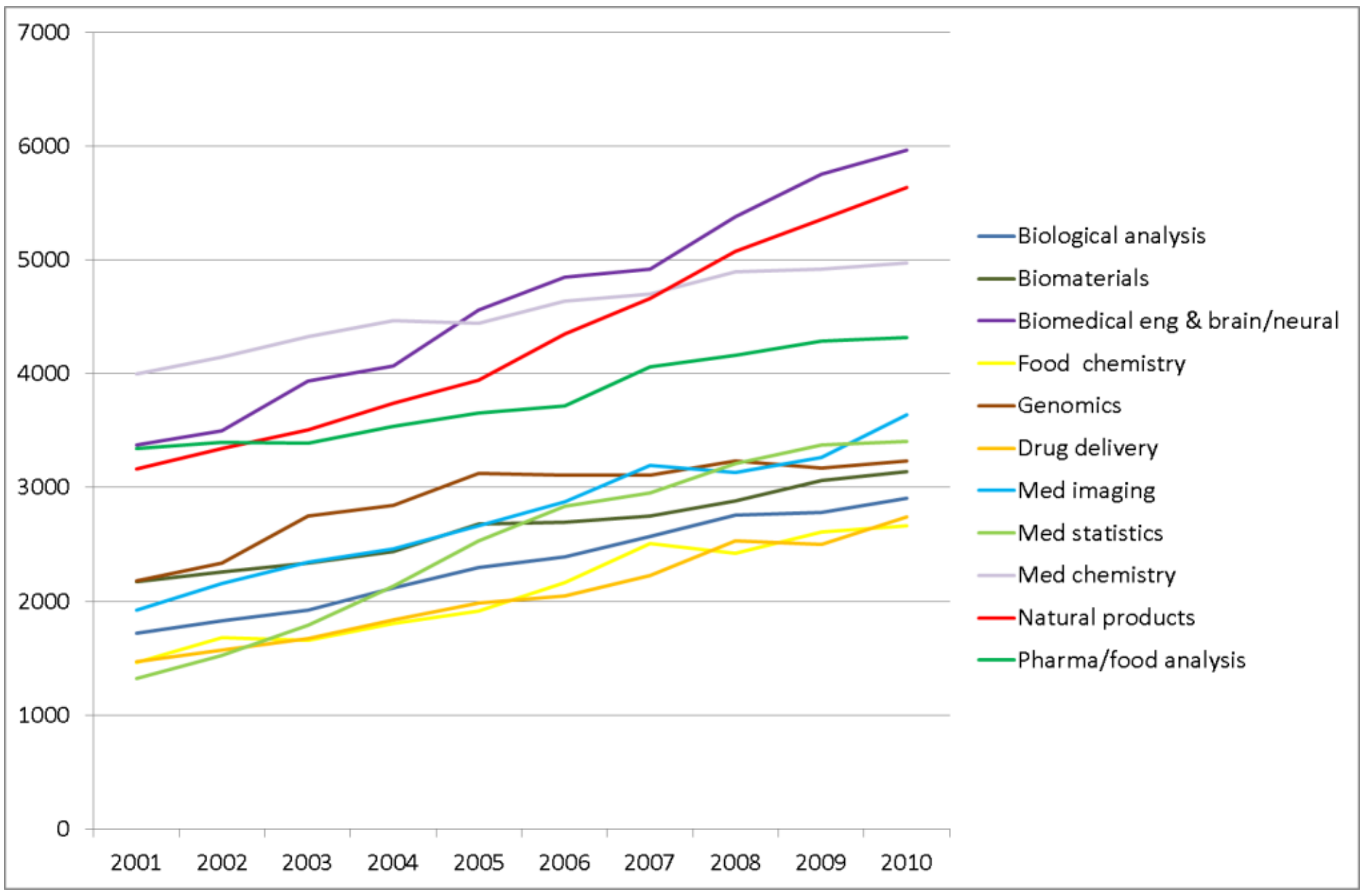

Figure 12. Yearly number of publications within each of the 11 broad research themes at the EPS-HLS interface.

Nevertheless, there turn out to be quite significant differences among the 11 broad research themes in their growth rates. Overall, the number of publications within the 11 research themes in 2010 is 63% larger than in 2001. However, there are two research themes with a much lower growth rate. These are *Medicinal chemistry* and *Pharmaceutical and food analysis*, with growth rates of 24% and 29%, respectively. On the other hand, there is one research theme with a growth rate that is about 2.5 times as high as the overall growth rate of 63%. This is *Medical statistics and informatics*. The number of publications within this theme in 2010 is 158% larger than in 2001. Such a high growth rate could potentially be an artifact of the database on which the analysis is based. For instance, as discussed above, it could be that the high growth rate is due to certain journals being added to the database during the period 2001–2010. However, we did not find any evidence of such database artifacts. We therefore conclude that the high growth rate of the *Medical statistics and informatics* theme is a genuine effect. Given the enormous increase in computer power and data availability, this high growth rate is very well understandable.

We emphasize that despite its high growth rate the *Medical statistics and informatics* theme is still among the smaller themes, in terms of its number of publications in 2010. As can be seen in Figure 12, the largest theme in 2010, *Biomedical engineering and brain/neural*, is almost twice as large. The other way around, the themes with the lowest growth rates, *Medicinal chemistry* and *Pharmaceutical and food analysis*, are still among the larger themes in terms of their number of publications in 2010.

A term map of the *Medical statistics and informatics* theme is presented in Figure 13. Colors are used to indicate the average age of the publications in which a term occurs. (So unlike in Figure 1 colors do not indicate citation impact.) Blue terms occur mainly in publications from the beginning of our period of analysis (2001–2010), while red terms occur mainly in publications from the most recent years. The term map suggests that the high growth rate of the *Medical statistics and informatics* theme is mainly due to bioinformatics research associated with proteomics and metabolomics, which is precisely the research area in which increases in computer power play a decisive role. This research area is shown in the upper-left part of the term map, in which many terms are colored orange or red.

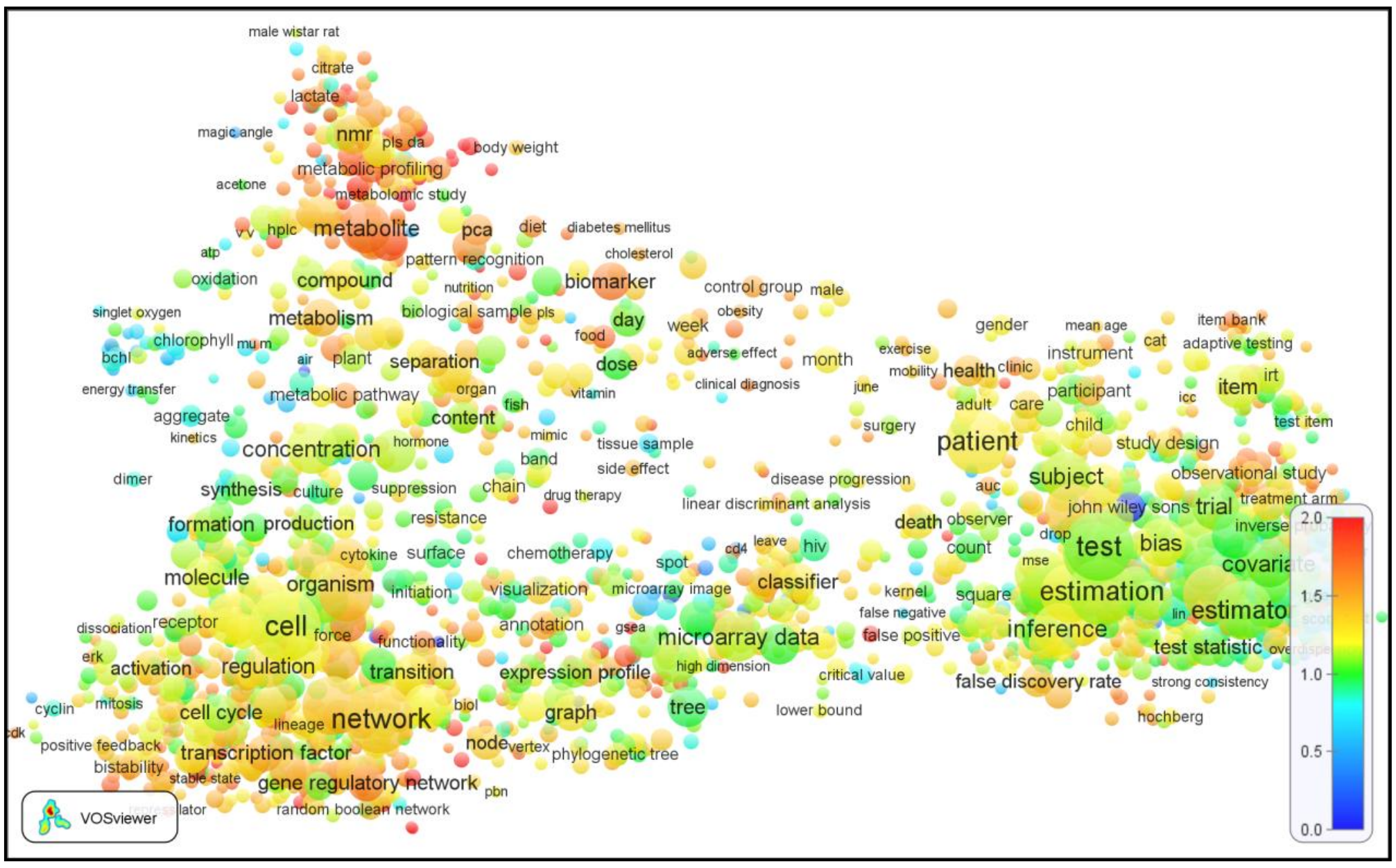

Figure 13. Term map of the *Medical statistics and informatics* research theme. Colors indicate the average age of the publications in which a term occurs.

*Contribution of countries, institutions, or research programs to research at the EPS-HLS interface*

Finally, our analysis can also provide information on the contribution of countries, institutions, or research programs of funding agencies to research at the EPS-HLS interface, particularly in terms of publication output and citation impact. In this sense, it also deals with the relation between interdisciplinarity and citation impact (e.g., Larivière & Gingras, 2010; Rafols, Leydesdorff, O'Hare, Nightingale, & Stirling, 2012).

As an example, we look at the contribution of the UK to each of the 11 broad research themes at the EPS-HLS interface. For each of the 11 research themes, we have calculated the percentage of the publications within the theme that have been co-authored by one or more UK research institutions. The results are reported in Table 6. As can be seen in the table, the UK has contributed most, in terms of the percentage of publications it has co-authored, to the *Medical statistics and informatics* research theme, which as discussed above is also the fastest growing theme. UK research institutions have co-authored 10.9% of the publications within this theme. Other research themes with a large UK contribution are *Biomedical engineering and brain/neural* and *Genomics and proteomics*, with respectively 9.3% and 9.1% of the publications within these themes being co-authored by UK research institutions. The research theme to which the UK has made the smallest contribution is the *Natural products for pharmaceutical use* theme. We find that 3.4% of the publications within this theme have been co-authored by UK research institutions. It should be noted that in some research themes publications co-authored by many different institutions from different countries may be more common than in other themes. This could partly account for the differences between research themes reported in Table 6.

Table 6. Contribution of the UK to the 11 research themes at the EPS-HLS interface.

| Research theme | % UK publications | UK citation impact score |
|---|---:|---:|
| Biological analysis | 5.6% | 0.96 |
| Biomaterials | 7.9% | 1.08 |
| Biomedical engineering and brain/neural | 9.3% | 1.12 |
| Food chemistry | 5.3% | 1.15 |
| Genomics and proteomics | 9.1% | 1.57 |
| Materials for drug delivery and controlled release | 5.6% | 1.12 |
| Medical imaging and radiotherapy | 7.2% | 1.35 |
| Medical statistics and informatics | 10.9% | 1.59 |
| Medicinal chemistry | 7.3% | 1.81 |
| Natural products for pharmaceutical use | 3.4% | 1.70 |
| Pharmaceutical and food analysis | 5.5% | 1.50 |

In addition to UK publication output, Table 6 also reports the citation impact of UK publications. Citation impact has been calculated as follows. For each research theme, we have determined the 10% most frequently cited publications. Next, we have determined the share of UK publications that are among the 10% most frequently cited within a research theme. For each research theme, the UK citation impact score equals the percentage of UK publications that are frequently cited divided by the overall percentage of frequently cited publications (which by definition is 10%). Thus, a citation impact score above one means that UK publications are performing above average in terms of citation impact. As can be seen in Table 6, with the exception of the *Biological analysis* research theme, the UK citation impact score is above one in all research themes. It is highest in the *Medicinal chemistry* theme, in which the UK's share of frequently cited publications is 81% above average. Other fields with a high UK citation impact score are *Natural products for pharmaceutical use* (1.70), *Medical statistics and informatics* (1.59), and *Genomics and proteomics* (1.57). Based on the results in Table 6, it is clear that the UK contributes significantly to high-impact research at the interface between EPS and HLS research fields.

## 4. Conclusions

The analysis presented in this report combines two different methodological approaches, on the one hand a textual approach based on term map visualizations and on the other hand a citation-based approach focusing on citation relations between publications. The textual approach directly considers the contents of publications, and therefore is very effective in providing many concrete examples of the influence of EPS research on HLS research. The citation-based approach, on the other hand, makes it possible to indicate, with a reasonable degree of accuracy, which publications in the scientific literature can be considered to be at the interface between EPS and HLS research fields. The strength of the citation-based approach is in revealing the structure of the scientific literature, both at the low level of individual research topics and at the higher level of broad research themes. Compared with the textual approach, the citation-based approach requires less human judgment and therefore is less sensitive to human subjectivity. The textual approach and the citation-based approach clearly have different strengths, and the two approaches can therefore be regarded as strongly complementary to each other.

With respect to the research question of the degree to which HLS advances are dependent on EPS research, the main results of our analysis can be summarized as follows:

- The dependence of HLS research on EPS research is visible in all 21 HLS fields that have been analyzed in detail using our textual approach. Looking at important terms occurring in the titles and abstracts of publications, it turns out that between 3% and 40% of the terms in an HLS field are directly related to EPS research.
- Some HLS fields can even be considered to be EPS-driven. An example of a clinical field for which this is the case is dentistry, which is strongly dependent on materials science research. In the life sciences, biomedical engineering is an example. Publications in HLS fields containing EPS-related terms in their titles and abstracts also often turn out to have an above-average citation impact.
- Our textual analysis reveals five major EPS topics that play a prominent role in HLS research: (1) new materials and their properties; (2) chemical methods for analysis and molecular synthesis; (3) imaging of parts of the body as well as of biomaterial surfaces; (4) medical engineering mainly related to imaging, radiation therapy, signal processing technology, and other medical instrumentation; and (5) mathematical and statistical methods for data analysis.
- Of all EPS and HLS publications, about 10% relates to topics that can be considered to be at the interface between EPS and HLS research fields. During the past decade, no increasing or decreasing trend could be detected in this percentage.
- Of the 11 broad research themes at the EPS-HLS interface that have been identified, the *Medical statistics and informatics* theme has by far experienced the largest growth in publication output during the past decade. The growth rate of this research theme has been 2.5 times above average. This appears to be mainly due to bioinformatics research associated with proteomics and metabolomics. Increasing computer power seems to play an essential role in this development.

- In terms of publication output, the UK is an important contributor to most research themes at the EPS-HLS interface. In terms of citation impact, in most research themes UK publications perform quite substantially above the worldwide average level.

Some of the above results have been obtained using our textual approach, others using the citation-based approach. The textual approach has given various detailed insights into the way in which EPS and HLS research interact with each other. The citation-based approach has provided a more high-level overview of EPS-HLS interaction along with various quantitative statistics on the characteristics of this interaction. We have also looked at the degree to which the two approaches have converged to similar results. Given the differences between the two approaches, checking for convergence turned out to be somewhat difficult. Nevertheless, we did find evidence of convergence. For instance, comparing the number of EPS terms in an HLS research field with the number of publications in the field that belong to the 11 broad research themes at the EPS-HLS interface, we found that fields such as dentistry and oncology have high scores on both dimensions. Likewise, fields with low score on one dimension typically also have a low score on the other dimension.

Various extensions of the analysis presented in this paper are possible. For instance, in the textual approach, the role played by EPS research in HLS fields could be investigated in more detail by making a classification of EPS-related terms into a number of different types (e.g., materials-related, technical, statistical, etc.). In the citation-based approach, a challenging extension would be to search for citation patterns that provide evidence of structural knowledge flows between fields, or perhaps even between series of fields, for instance from physics to chemistry to the life sciences to medicine. Another possible extension would be to systematically monitor how different fields of science depend on each other, how these dependencies evolve over time, and how they influence the emergence of new interdisciplinary research areas. Within this context, the contribution made by different countries to research areas at the boundary between disciplines could be monitored as well, and with the improving availability of funding data in bibliographic databases, also the role played by individual funding agencies could be analyzed.

## Acknowledgments

We thank Nees Jan van Eck (CWTS, Leiden University) for his help in creating the term maps used in Section 2. We thank Qi Wang (KTH Royal Institute of Technology) for her contribution to the development of the approach used in Subsection 3.2 for identifying emerging topics.

## Appendix

Table A1. The 72 EPS research fields (WoS journal subject categories) in the analysis discussed in Section 3.

| | |
|---|---|
| acoustics | materials science, ceramics |
| astronomy & astrophysics | materials science, characterization & testing |
| automation & control systems | materials science, coatings & films |
| biophysics | materials science, composites |
| chemistry, analytical | materials science, multidisciplinary |
| chemistry, applied | materials science, paper & wood |
| chemistry, inorganic & nuclear | materials science, textiles |
| chemistry, medicinal | mathematical & computational biology |
| chemistry, multidisciplinary | mathematics |
| chemistry, organic | mathematics, applied |
| chemistry, physical | mathematics, interdisciplinary applications |
| computer science, artificial intelligence | mechanics |
| computer science, cybernetics | metallurgy & metallurgical engineering |
| computer science, hardware & architecture | microscopy |
| computer science, information systems | mining & mineral processing |
| computer science, interdisciplinary applications | nanoscience & nanotechnology |
| computer science, software engineering | nuclear science & technology |
| computer science, theory & methods | operations research & management science |
| construction & building technology | optics |
| crystallography | physics, applied |
| electrochemistry | physics, atomic, molecular & chemical |
| energy & fuels | physics, condensed matter |
| engineering, aerospace | physics, fluids & plasmas |
| engineering, biomedical | physics, mathematical |
| engineering, chemical | physics, multidisciplinary |
| engineering, civil | physics, nuclear |
| engineering, electrical & electronic | physics, particles & fields |
| engineering, industrial | polymer science |
| engineering, manufacturing | robotics |
| engineering, mechanical | social sciences, mathematical methods |
| engineering, multidisciplinary | spectroscopy |
| engineering, petroleum | statistics & probability |
| ergonomics | telecommunications |
| instruments & instrumentation | thermodynamics |
| logic | transportation |
| materials science, biomaterials | transportation science & technology |

Table A2. The 86 HLS research fields (WoS journal subject categories) in the analysis discussed in Section 3.

| | |
|---|---|
| agricultural engineering | medical informatics |
| agricultural experiment station reports | medical laboratory technology |
| agriculture, dairy & animal science | medicine, general & internal |
| agriculture, multidisciplinary | medicine, research & experimental |
| agronomy | microbiology |
| allergy | mycology |
| anatomy & morphology | neuroimaging |
| andrology | neurosciences |
| anesthesiology | nursing |
| audiology & speech-language pathology | nutrition & dietetics |
| behavioral sciences | obstetrics & gynecology |
| biochemical research methods | oncology |
| biochemistry & molecular biology | ophthalmology |
| biology | ornithology |
| biophysics | orthopedics |
| biotechnology & applied microbiology | otorhinolaryngology |
| cardiac & cardiovascular systems | parasitology |
| cell & tissue engineering | pathology |
| cell biology | pediatrics |
| clinical neurology | peripheral vascular disease |
| critical care medicine | pharmacology & pharmacy |
| dentistry/oral surgery & medicine | physiology |
| dermatology | plant sciences |
| developmental biology | primary health care |
| emergency medicine | psychiatry |
| endocrinology & metabolism | public, environmental & occupational health |
| entomology | radiology, nuclear medicine & medical imaging |
| evolutionary biology | rehabilitation |
| fisheries | reproductive biology |
| food science & technology | respiratory system |
| gastroenterology & hepatology | rheumatology |
| genetics & heredity | social work |
| geriatrics & gerontology | soil science |
| gerontology | sport sciences |
| health care sciences & services | substance abuse |
| health policy & services | surgery |
| hematology | toxicology |
| horticulture | transplantation |
| immunology | tropical medicine |
| infectious diseases | urology & nephrology |
| integrative & complementary medicine | veterinary sciences |
| marine & freshwater biology | virology |
| mathematical & computational biology | zoology |